\documentstyle[11pt,twoside]{article}
\newcommand{\beq}{\begin{equation}}
\newcommand{\eeq}{\end{equation}}
\newcommand{\beqa}{\begin{eqnarray}}
\newcommand{\eeqa}{\end{eqnarray}}
\newcommand{\beqan}{\begin{eqnarray*}}
\newcommand{\eeqan}{\end{eqnarray*}}
\newcommand{\ba}{\begin{array}}
\newcommand{\ea}{\end{array}}
\newcommand{\no}{\nonumber}

\newcommand{\ra}{\rightarrow}

\newcommand{\ve}{\varepsilon}

\newcommand{\dg}{\dagger}

\newcommand{\wh}{\widehat}

\newcommand{\A}{{\cal A}}

\newcommand{\D}{{\cal D}}

\newcommand{\cL}{{\cal L}}
\newcommand{\M}{{\cal M}}

\newcommand{\V}{{\cal V}}

\newcommand{\dfrac}{\displaystyle \frac}

\newcommand{\dlim}{\displaystyle \lim}

\normalsize
\begin{document}
\begin{titlepage}
\begin{flushright}
UWThPh-1994-1\\
January 1994\\
hep-ph/9402337
\end{flushright}
\vspace{3cm}
\begin{center}
{\Large \bf Chiral Invariant Renormalization of the \\[5pt]
Pion--Nucleon Interaction*}\\[40pt]
G. Ecker  \\
Institut f\"ur Theoretische Physik \\
Universit\"at Wien \\
Boltzmanngasse 5, A-1090 Vienna, Austria
\vfill
{\bf Abstract} \\
\end{center}
\noindent
The leading divergences of the generating functional for Green
functions of quark currents between one--nucleon states are
calculated with heat kernel techniques. The results allow for
a chiral invariant renormalization of all two--nucleon Green
functions of the pion--nucleon system to $O(p^3)$ in the
low--energy expansion.

\vfill
\noindent * Work supported in part by FWF, Project No. P09505--PHY and
by HCM, EEC--Contract No. CHRX--CT920026 (EURODA$\Phi$NE)
\end{titlepage}

\paragraph{1.} The modern treatment of the pion--nucleon interaction as
an effective field theory of the standard model was pioneered by Gasser,
Sainio and \v{S}varc \cite{GSS} who applied the methods of
chiral perturbation theory (CHPT) to the $\pi N$ system. However,
the chiral expansion with baryons is hampered by the presence
of the nucleon mass, which stays finite in the chiral limit. It was
then shown by Jenkins and Manohar \cite{JM1} that the methods of
heavy quark effective theory \cite{HQET} allow for a systematic
low--energy expansion of baryonic Green functions in complete
analogy to the meson sector.

The purpose of this letter is to perform the complete renormalization of
Green functions of quark currents between one--nucleon states to
$O(p^3)$ in the chiral expansion. This requires
a chiral invariant calculation of the divergent part of the corresponding
generating functional at the one--loop level. Many specific Green
functions of the $\pi N$ system have already been analyzed
to one--loop accuracy \cite{GSS,JM2,RMei,Rho}. However, only the full
divergence structure of $O(p^3)$ presented here permits a complete
renormalization of all Green functions: nucleon form factors,
$\pi N \ra \pi \ldots \pi N$, $\gamma^* N \ra \pi \ldots \pi N$,
$W^* N \ra \pi \ldots \pi N$, etc. Results will be given for two
light flavours only. All details and the extension to chiral $SU(3)$ will
be presented elsewhere.

We start from QCD with two light flavours $u,d$
coupled to external hermitian fields \cite{GLAP}:
\beq
\cL = \cL^0_{\rm QCD} + \bar q \gamma^\mu \left(\V_\mu + \frac{1}{3}
\V^s_\mu + \gamma_5 \A_\mu\right) q - \bar q (S - i\gamma_5 P)q, \qquad
q = \left( \ba{c} u \\ d \ea \right). \label{QCD}
\eeq
$S,P$ are general 2--dimensional matrix fields, the isotriplet vector
and axial--vector fields $\V_\mu,\A_\mu$ are traceless and the
isosinglet vector field $\V^s_\mu$ is included to generate the
electromagnetic current. At the effective level of pions and nucleons,
$\V^s$ couples directly only to nucleons since the pions have
zero baryon number.

Explicit chiral symmetry breaking is implemented by setting
$S = \M = \mbox{diag }(m_u,m_d)$.
The chiral group $G = SU(2)_L \times SU(2)_R$ breaks spontaneously to
$SU(2)_V$ (isospin). It is realized non--linearly \cite{CCWZ}
on the Goldstone pion fields $\phi$:
\beqa
\xi_L(\phi) & \stackrel{g}{\ra} & g_L \xi_L(\phi) h(g,\phi)^{-1}, \qquad
g = (g_L,g_R) \in G \no \\
\xi_R(\phi) & \stackrel{g}{\ra} & g_R \xi_R(\phi) h(g,\phi)^{-1}, \label{coset}
\eeqa
where $\xi_L,\xi_R$ are elements of the chiral coset space
$SU(2)_L \times SU(2)_R/SU(2)_V$ and the compensator field
$h(g,\phi)$ is in $SU(2)_V$.
The more familiar quantity $U(\phi) = \xi_R(\phi) \xi_L(\phi)^\dg$
transforms linearly under $G$. In the
standard ``gauge'' with $\xi_R = \xi_L^\dg =: u$ we have $U = u^2$.

The nucleon doublet $\Psi$ transforms as
\beq
\Psi = \left( \ba{c} p \\ n \ea \right) \stackrel{g}{\ra}
\Psi' = h(g,\phi)\Psi \label{psi}
\eeq
under chiral transformations. The local nature of this transformation
requires a connection
\beq
\Gamma_\mu = \frac{1}{2} \{ \xi_R^\dg (\partial_\mu - i r_\mu)\xi_R +
\xi_L^\dg (\partial_\mu - i \ell_\mu)\xi_L\} \label{conn}
\eeq
in the presence of external gauge fields
\beq
r_\mu = \V_\mu + \A_\mu , \qquad \ell_\mu = \V_\mu - \A_\mu
\label{gf}
\eeq
to define a covariant derivative
\beq
\nabla_\mu \Psi = (\partial_\mu + \Gamma_\mu - i \V^s_\mu)\Psi .
\eeq

The effective chiral Lagrangian for Green
functions with at most two nucleons is \cite{GSS,GLAP}
\beqa
\cL_{\rm eff} &=& \cL_M + \cL_{MB} \label{LMB} \\
\cL_M &=& \cL_2 + \cL_4 + \ldots \no \\
\cL_2 &=& \frac{F^2}{4} \langle D_\mu U D^\mu U^\dg + \chi U^\dg +
\chi^\dg U \rangle = \frac{F^2}{4} \langle u_\mu u^\mu + \chi_+ \rangle
\label{L2} \\
&& u_\mu = i \{ \xi_R^\dg (\partial_\mu - i r_\mu)\xi_R -
\xi_L^\dg (\partial_\mu - i \ell_\mu) \xi_L\} \no \\
&& \chi = 2B_0(S + iP), \qquad \chi_\pm = u^\dg \chi u^\dg
\pm u \chi^\dg u \no \\
\cL_{MB} &=& \cL_{\pi N}^{(1)} + \cL_{\pi N}^{(2)} + \cL_{\pi N}^{(3)}
+ \ldots \no \\
\cL_{\pi N}^{(1)} &=& \bar \Psi (i \not\!\nabla - m + \frac{g_A}{2}
\not\!u \gamma_5) \Psi \label{piN1}
\eeqa
where $m,g_A$ are the nucleon mass and the neutron decay constant in the
chiral limit and $\langle \dots \rangle$ stands for the trace in
flavour space.

The generating functional of Green functions $Z[j,\eta,\bar \eta]$ is
defined \cite{GSS} by the path integral
\beq
e^{iZ[j,\eta,\bar\eta]} = N \int [du d\Psi d \bar \Psi]
\exp [i\{ S_M + S_{MB} + \int d^4 x (\bar \eta \Psi + \bar \Psi \eta)\}].
\label{Z}
\eeq
The action $S_M + S_{MB}$ corresponds to the effective Lagrangian
(\ref{LMB}), the external fields $\V,\A,S,P$ are denoted collectively
as $j$ and $\eta,\bar \eta$ are fermionic sources.

\paragraph{2.} Heavy baryon CHPT \cite{JM1} can be viewed as a clever
choice of variables for performing the fermionic path integral in
(\ref{Z}). By shifting the dependence on the nucleon mass $m$ from
the nucleon propagator to the vertices of the effective Lagrangian,
the integration over the new fermionic variables produces a systematic
low--energy expansion.

In terms of the velocity--dependent fields $N_v,H_v$ defined as
\beqa
N_v(x) &=& \exp[i m v \cdot x] P_v^+ \Psi(x) \label{vdf} \\
H_v(x) &=& \exp[i m v \cdot x] P_v^- \Psi(x) \no \\
P_v^\pm &=& \frac{1}{2} (1 \pm \not\!v)~, \qquad v^2 = 1 ~, \no
\eeqa
the pion--nucleon action $S_{MB}$ takes the form
\beqa
S_{MB} &=& \int d^4 x \{ \bar N_v A N_v + \bar H_v B N_v +
\bar N_v \gamma^0 B^\dg \gamma^0 H_v - \bar H_v C H_v\} \label{SMB} \\
A &=& iv \cdot \nabla + g_A S \cdot u + A_{(2)} + A_{(3)} + \ldots \no \\
B &=& i \not\!\nabla^\perp - \frac{g_A}{2} v \cdot u \gamma_5 +
B_{(2)} + B_{(3)} + \ldots \no \\
C &=& 2m + i v \cdot \nabla + g_A S \cdot u + C_{(2)} + C_{(3)} + \ldots
\no \\
&& \nabla_\mu^\perp = \nabla_\mu - v_\mu v \cdot \nabla, \qquad
[\not\!v,A] = [\not\!v,C] = 0, \qquad \{\not\!v,B\} = 0. \no
\eeqa
In $A$ and $C$, the only dependence on Dirac matrices is through
the spin matrix
\beq
S^\mu = \frac{i}{2} \gamma_5 \sigma^{\mu\nu} v_\nu , \qquad
S \cdot v = 0, \qquad S^2 = - \frac{3}{4} {\bf 1}.
\eeq

Rewriting also the source term in (\ref{Z}) in terms of
$N_v,H_v$ with corresponding sources
\beq
\rho_v = e^{imv \cdot x} P_v^+ \eta , \qquad
R_v = e^{imv \cdot x} P_v^- \eta ~,
\eeq
one can now integrate out the ``heavy'' components $H_v$
 to obtain a non--local action in the ``light''
fields $N_v$ \cite{MRR,BKKM,EPrag}. At this point, the crucial
approximation of heavy baryon CHPT is made:
the action is written as a series of local actions with
increasing chiral dimensions by expanding $C^{-1}$ in a power series
in $1/m$:
\beq
C^{-1} = \frac{1}{2m} - \frac{i v \cdot \nabla + g_A S \cdot u}{(2m)^2}
+ O(p^2)~.
\eeq
With this approximation, the integration over $N_v$ reduces again to
completing a square (the fermion determinant is trivial to any
finite order in $1/m$) with the final result
\beq
e^{iZ[j,\eta,\bar\eta]} = N \int [du]
e^{i(S_M + Z_{MB}[u,j,\rho_v,R_v])} \label{mfi}
\eeq
where
\beqa
\lefteqn{Z_{MB}[u,j,\rho_v,R_v] = - \int d^4 x \{ \bar \rho_v
(A + \gamma^0 B^\dg \gamma^0 C^{-1} B)^{-1} \rho_v} \no \\
&& \mbox{} + \bar R_v C^{-1} B(A + \gamma^0 B^\dg \gamma^0 C^{-1} B)^{-1}
\rho_v + \bar \rho_v (A + \gamma^0 B^\dg \gamma^0 C^{-1} B)^{-1}
\gamma^0 B^\dg \gamma^0 C^{-1} R_v \no \\
&& \mbox{} + \bar R_v C^{-1} B (A + \gamma^0 B^\dg \gamma^0 C^{-1} B)^{-1}
\gamma^0 B^\dg \gamma^0 C^{-1} R_v - \bar R_v C^{-1} R_v\}.
\label{ZMBU}
\eeqa
The functional integral (\ref{Z}) has been reduced to the mesonic integral
(\ref{mfi}). From here on, the standard procedure of CHPT \cite{GLAP}
can be applied: the action in the functional integral (\ref{mfi}) is
expanded around the classical solution $u_{\rm cl}[j]$ of the
lowest--order equation of motion. Integration over the quantum
fluctuations gives rise to a well--behaved low--energy expansion for
$Z[j,\eta,\bar \eta]$ like in the meson sector.

For $u = u_{\rm cl}$, we can read off the tree--level nucleon
propagator from the right--hand side of Eq. (\ref{ZMBU}) (see Fig.~1
of Ref. \cite{GSS} for an artistic impression of this object):
\beq
Z_{MB}^{\rm tree} [j,\eta,\bar \eta] = Z_{MB} [u_{\rm cl}[j],j,\rho_v,
R_v]. \label{tree}
\eeq
By construction, $Z_{MB}^{\rm tree}$ is
independent of the arbitrary vector $v$. It is instructive to check
this independence for the free propagator corresponding to
\footnote{$C^{-1}$ is not expanded in this case.}
\beq
A = iv \cdot \partial, \qquad B = i \not\!\partial^\perp, \qquad
C = iv \cdot \partial + 2m.
\eeq
Inserting these operators into the right--hand side of Eq. (\ref{ZMBU})
and reexpressing $\rho_v,R_v$ in terms of the original fermionic
sources, one finds indeed the free massive fermion propagator
\beq
Z_{MB}^{\rm tree}[0,\eta,\bar \eta] = \int d^4x \bar \eta(x)
(i \not\!\partial + m)(\Box + m^2)^{-1} \eta(x).
\eeq
On the other hand, any given order in the chiral expansion of
$Z_{MB}[j,\eta,\bar \eta]$ will in general not be independent of
$v$ because a change in $v$ involves
different chiral orders (reparametrization invariance \cite{LuMa}).
This is the price one has to pay for a systematic low--energy expansion.

\paragraph{3.} We now turn to the calculation of $Z_{MB}[j,\eta,\bar \eta]$
up to and including $O(p^3)$. Since we only consider nucleons (rather
than antinucleons), we can drop the sources $R_v$ to the order
considered. To $O(p^2)$, $Z_{MB}[j,\eta,\bar \eta]$ is a pure tree--level
functional. The relevant pion--nucleon Lagrangian is \cite{JM1,BKKM}
\beq
\bar N_v(A + \gamma^0 B^\dg \gamma^0 C^{-1} B)N_v =
\bar N_v \left( A_{(1)} + A_{(2)} + \frac{1}{2m} \gamma^0
B_{(1)}^\dg \gamma^0 B_{(1)}\right) N_v  + O(p^3)
\eeq
$$
A_{(1)} = iv \cdot \nabla + g_A S \cdot u  \qquad
B_{(1)} = i \not\!\nabla^\perp - \frac{g_A}{2} v \cdot u \gamma_5
$$
\beqa
A_{(2)} &=& C_1 \langle u \cdot u \rangle +
C_2 \langle (v \cdot u)^2\rangle + C_3 \chi_+ + C_4 \langle \chi_+\rangle
\label{A2} \\
&& \mbox{} + \ve^{\mu\nu\rho\sigma} v_\rho S_\sigma \{ C_5 u_\mu u_\nu +
C_6 f_{+\mu\nu} + C_7 (\partial_\mu \V^s_\nu - \partial_\nu \V^s_\mu)\}
\no \\
f_\pm^{\mu\nu} &=& u F_L^{\mu\nu} u^\dg \pm u^\dg F_R^{\mu\nu} u, \no
\eeqa
where $F_{L,R}$ are the field strengths associated with the isotriplet
gauge fields $\ell,r$ in (\ref{gf}). Since the loop contributions set in
at $O(p^3)$ only, neither $g_A$ nor the coupling constants $C_1,\ldots,
C_7$ of $O(p^2)$ will have to be renormalized.

To calculate the loop functional of $O(p^3)$, we expand
\beq
\cL_2 + \cL_4 - \bar \rho_v A_{(1)}^{-1} \rho_v
\eeq
in the functional integral (\ref{mfi}) around the classical solution
$u_{\rm cl}[j]$. A convenient choice of fluctuation variables $\xi$ is
given by \cite{GLAP}
\beq
\ell_R(\phi) = u(\phi_{\rm cl})e^{i\xi(\phi)/2}, \qquad
\ell_L(\phi) = u^\dg(\phi_{\rm cl})e^{-i\xi(\phi)/2}, \qquad
\xi^\dg = \xi, \qquad \langle \xi \rangle = 0, \qquad \xi(\phi_{\rm cl})
= 0.
\eeq
Following Gasser, Sainio and \v{S}varc \cite{GSS}, we expand $\cL_2$ to
$O(\xi^3)$, $\cL_4$ to $O(\xi)$ and $A^{(1)}$ to $O(\xi^2)$ to arrive
at the diagrams of Figs.~1,2. The sum of the reducible diagrams in
Fig.~2 is finite and scale independent \cite{GLAP}. For the
irreducible diagrams of Fig.~1, we need
\beqa
A_{(1)} &=& iv \cdot \nabla + g_A S \cdot u  \no \\
&=& iv \cdot \nabla_{\rm cl} + g_A S \cdot u_{\rm cl}
+ \frac{i}{4} [v \cdot u_{\rm cl},\xi] - g_A S \cdot \nabla_{\rm cl} \xi
\no \\
&& \mbox{} + \frac{i}{8} \xi v \cdot
\stackrel{\leftrightarrow}{\nabla}_{\rm cl} \xi +
\frac{g_A}{8} [\xi,[S \cdot u_{\rm cl},\xi]] + O(\xi^3) \\
\nabla^\mu_{\rm cl} \xi &:=& \partial^\mu \xi + [\Gamma^\mu_{\rm cl},\xi].
\no
\eeqa
 From now on, the index cl will be dropped. All mesonic quantities
are to be taken at the classical solution of the lowest--order equation
of motion.

The diagrams of Fig. 1 correspond to the following generating functional:
\beq
Z_{\rm irr}[j,\rho_v] = \int d^4x d^4x' d^4y d^4y' \bar\rho_v(x)
A^{-1}_{(1)}(x,y) [\Sigma_1(y,y') \delta^4(y-y') + \Sigma_2(y,y')] \cdot
A^{-1}_{(1)}(y',x') \rho_v(x') \label{Zirr}
\eeq
where $A^{-1}_{(1)}$ is the propagator for $N_v$ in the presence
of external fields. The self--energy functionals $\Sigma_1,\Sigma_2$
are given by
\beqa
\Sigma_1(y,y') &=& \frac{1}{8F^2} \{i \tau_i [G_{ij}(y,y') v \cdot
\stackrel{\leftarrow}{d'}_{jk} - v \cdot d_{ij} G_{jk}(y,y')]\tau_k
+ g_A [\tau_i,[S \cdot u,\tau_j]] G_{ij}(y,y')\} \label{Si1} \\
\Sigma_2(y,y') &=& - \frac{2}{F^2} V_i(y) G_{ij}(y,y')
A^{-1}_{(1)}(y,y') V_j (y') \label{Si2} \\
V_i &=& \frac{i}{4 \sqrt{2}} [v \cdot u,\tau_i] - \frac{g_A}{\sqrt{2}}
\tau_j S \cdot d_{ji} . \no
\eeqa
The differential operator $d^\mu_{ij}$ in 3--dimensional tangent space is
related to the previously defined covariant derivative as
\beq
\nabla^\mu \xi = \frac{1}{\sqrt{2}} \tau_i  d^\mu_{ij}\xi_j , \qquad
\xi = \frac{1}{\sqrt{2}} \tau_i \xi_i , \qquad
d^\mu_{ij} = \delta_{ij} \partial^\mu + \gamma^\mu_{ij}, \qquad
\gamma^\mu_{ij} = - \frac{1}{2} \langle \Gamma^\mu [\tau_i,\tau_j]\rangle.
\eeq
It acts on the meson propagator $G_{ij}$:
\beq
G = (d_\mu d^\mu + \sigma)^{-1} , \qquad
\sigma_{ij} = \frac{1}{8} \langle [u_\mu,\tau_i][\tau_j,u^\mu] +
\chi_+ \{ \tau_i,\tau_j\} \rangle.
\eeq

\paragraph{4.} The self--energy functionals $\Sigma_1(y,y')$,
$\Sigma_2(y,y')$ are divergent for $y' \ra y$. In order to extract the
divergences in a chiral invariant manner, we use the heat kernel
representation of propagators (see Ref. \cite{Ball} for a review) in
$d$--dimensional Euclidean space. The divergences will appear as simple
poles in $\ve = \frac{1}{2} (4-d)$. The corresponding residua are local
polynomials in the fields of $O(p^3)$ and can easily be transformed
back to Minkowski space.

In Euclidean space with $d$ dimensions, the inverse of the elliptic
second--order differential operator
\beq
D_2 = - d_\mu d_\mu + \sigma
\eeq
can be constructed as an integral
\beq
G(x,y) = D^{-1}_2(x,y) = \int_0^\infty d\lambda G(x,y;\lambda)
\eeq
over the heat kernel $G(x,y;\lambda)$ with its asymptotic expansion for
$\lambda \ra 0_+$~:
\beqa
&& G(x,y;\lambda) = \frac{1}{(4\pi\lambda)^{d/2}} e^{- (x-y)^2/4\lambda}
\sum_n a_n(x,y) \lambda^n \no \\
&& \left( \frac{\partial}{\partial \lambda} + D_2\right) G(x,y;\lambda)
= 0, \qquad G(x,y;0) = \delta^d(x-y).
\eeqa
The differential equation for $G(x,y;\lambda)$ yields recursion relations
for the Seeley--DeWitt coefficients $a_n$, which are matrices in
tangent space of $O(p^{2n})$.
Since the divergences of $\Sigma_1,\Sigma_2$ are $O(p^3)$, we only need
the following derivatives of $a_0,a_1$ in the coincidence limit $x\ra y$,
which is to be understood in the list below:
\beq
\ba{ll}
a_0 = {\bf 1} &  \\[10pt]
d_\mu a_0 = a_0 \stackrel{\leftarrow}{d_\mu}\; = 0 &
d_\mu a_n = \dlim_{x \ra y} [\partial^x_\mu + \gamma_\mu(x)] a_n(x,y) \\[10pt]
d_\mu d_\nu a_0 = d_\mu a_0 \stackrel{\leftarrow}{d_\nu} \;= \dfrac{1}{2}
\gamma_{\mu\nu} &
a_n \stackrel{\leftarrow}{d_\mu}\; = \dlim_{x \ra y} [\partial^y_\mu
a_n(x,y) - a_n(x,y) \gamma_\mu(y)] \\[10pt]
d_\lambda d_\mu d_\nu a_0 = 2d_\lambda d_\mu a_0 \stackrel{\leftarrow}
{d_\nu}\; = \dfrac{1}{3}(d_\lambda \gamma_{\mu\nu} +
d_\mu \gamma_{\lambda\nu})
& \gamma_{\mu\nu} = \partial_\mu \gamma_\nu - \partial_\nu \gamma_\mu +
[\gamma_\mu, \gamma_\nu] \\[10pt]
a_1 = - \sigma
 & d_\lambda \gamma_{\mu\nu} = \partial_\lambda \gamma_{\mu\nu} +
[\gamma_\lambda,\gamma_{\mu\nu}] \\[10pt]
d_\mu a_1 = \dfrac{1}{6} d_\nu \gamma_{\mu\nu} - \dfrac{1}{2} d_\mu \sigma
& \\[10pt]
a_1 \stackrel{\leftarrow}{d_\mu} \; = - \dfrac{1}{6} d_\nu \gamma_{\mu\nu}
- \dfrac{1}{2} d_\mu \sigma.
\ea
\eeq
In $d$ dimensions, we have \cite{Ball,JO}
\beqa
&& G(x,x) = \frac{a_1(x,x)}{(4\pi)^2 \ve}, \qquad
\ve = \frac{4 - d}{2} \no \\
&& \lim_{x \ra y} \partial^x_\mu (\partial^y_\mu)G(x,y) =
\frac{1}{(4\pi)^2\ve} \lim_{x\ra y} \partial^x_\mu (\partial^y_\mu)
a_1 (x,y).
\eeqa
Inserting the coincidence relations for $a_1$ into $\Sigma_1$ in
(\ref{Si1}), performing the summation over the tangent space indices
$i,j,k$ and transforming back to Minkowski space, one obtains the
divergent part of $\Sigma_1$, which is manifestly of $O(p^3)$:
\beqa
\Sigma_1^{\rm div}(y,y) &=& \frac{1}{(4\pi F)^2\ve} \wh \Sigma_1(y)
\label{Si1div} \\
\wh \Sigma_1(y) &=&  - \frac{i}{6} \nabla^\mu \Gamma_{\mu\nu} v^\nu
 + \frac{g_A}{8} \left( \{ S \cdot u, u \cdot u + \chi_+\}
+ S \cdot u \langle u \cdot u + \chi_+\rangle \right.\no \\
&& \left. \mbox{} + 2 u_\mu \langle u^\mu S \cdot u\rangle
 - \langle S \cdot u u \cdot u\rangle -
\langle S \cdot u \chi_+\rangle \right)  \\
\Gamma_{\mu\nu} &=& \partial_\mu \Gamma_\nu - \partial_\nu \Gamma_\mu +
[\Gamma_\mu,\Gamma_\nu] . \no
\eeqa

\paragraph{5.} The divergences of $\Sigma_2(y,y')$
are due to the singular behaviour for $y \ra y'$ of the
product of (derivatives of) propagators
\beq
G_{ij}(y,y') A^{-1}_{(1)} (y,y')~.
\eeq
In accordance with locality, the
divergences are of the generic form (with an overall chiral dimension 3)
\beq
\frac{1}{(4\pi F)^2\ve} F(y,y') D(y;v) \delta^4(y-y')
\eeq
with a field monomial $F(y,y')$ and a differential operator $D(y;v)$
of at most third order.

The missing ingredient is the heat kernel representation for the
nucleon propagator $A^{-1}_{(1)}$. Below, only results
needed for the present analysis are collected. However, the following
representation will clearly also be useful in heavy quark effective theory.

Since $A_{(1)}$ is not an elliptic differential operator, we write
\beq
A^{-1}_{(1)} = i \D^\dg (\D \D^\dg)^{-1} , \qquad \qquad \D = i A_{(1)}
\eeq
and set up a heat kernel representation for the inverse of the
positive definite hermitian operator $\Delta = \D \D^\dg$, where
\beqa
\D &=& - v \cdot \nabla - i g_A S \cdot u + \eta v^2, \qquad
\eta \ra 0_+ \no \\
\Delta &=& \D \D^\dg = - (v \cdot \nabla)^2 + R + \eta^2 \\
R &=& i g_A ( v \cdot \nabla S \cdot u) - g^2_A (S \cdot u)^2. \no
\eeqa
To facilitate the transformation back to Minkowski space, it is
preferable to keep track of the dependence on $v^2$ in Euclidean space.
To get a well--defined representation
for $\Delta^{-1}$ without infrared divergences, one must keep
$\eta \neq 0$ in intermediate steps. The heat
kernel representation for $\Delta^{-1}$ can be motivated
by remembering that $\Delta$ is essentially a one--dimensional
differential operator in the direction of $v$:
\beq
\Delta^{-1}(x,y) = \int_0^\infty d\tau \; E(x,y;\tau)
\eeq
$$
\left( \frac{\partial}{\partial \tau} + \Delta\right) E(x,y;\tau) = 0,
 \qquad E(x,y;0) = \delta^d(x-y)
$$
$$
E(x,y;\tau) = \frac{1}{\sqrt{4\pi\tau}} \exp \left[-
\frac{[v \cdot (x-y)]^2}{4\tau} - \eta^2\tau \right] g(x-y;v)
\sum_n b_n (x,y) \tau^n
$$
$$
g(x;v) = \int \frac{d^d k}{(2\pi)^{d-1}} \delta(k\cdot v) e^{-ik\cdot x},
\qquad v \cdot \partial g = 0.
$$
The differential equation for $E$ produces again recursion
relations for the coefficients $b_n$, which are matrices in 2--dimensional
flavour space. The following values of the $b_n(x,x)$
are needed for the present analysis:
\beqa
b_0 = {\bf 1}, && (v \cdot \nabla)^n b_0 = 0 \no \\
b_1 = - R, && v \cdot \nabla b_1 = - \frac{1}{2} v \cdot \nabla R.
\eeqa

In addition to the Seeley--DeWitt coefficients $a_n,b_n$ and their
derivatives, we need the coincidence limits of (derivatives of)
products of the meson and nucleon propagator functions
\beq
G_n(x) = \int_0^\infty d\lambda \frac{\lambda^n \exp[-\frac{x^2}{4\lambda}]}
{(4\pi \lambda)^{d/2}}, \qquad
E_m(x;v) = g(x;v) \int_0^\infty d\tau \frac{\tau^m
\exp[- \frac{(v \cdot x)^2}{4\tau} - \eta^2\tau]}{\sqrt{4 \pi \tau}}.
\eeq
The corresponding singularities can best be extracted
in Euclidean $d$-dimensional Fourier space \cite{JO}.
The following list contains
all singular products appearing in $\Sigma_2$:
\beqa
G_0(x) E_0(x;v) &\sim& - \frac{2}{(4\pi)^2v^2\ve} \delta^d(x)
\qquad \ve = \frac{1}{2}(4-d) \\
G_0(x)v \cdot \partial E_0(x;v) &\sim&  \frac{2}{(4\pi)^2v^2\ve}
v \cdot \partial \delta^d(x) \\
S_\mu S_\nu \partial_\mu \partial_\nu G_0(x) E_0(x;v) &\sim&
 \frac{2}{(4\pi)^2\ve} S_\mu S_\nu \delta_{\mu\nu} (v \cdot \partial)^2
\delta^d(x) \\
S_\mu S_\nu \partial_\mu \partial_\nu G_0(x)v \cdot \partial E_0(x;v) &\sim&
- \frac{2}{3(4\pi)^2\ve} S_\mu S_\nu \delta_{\mu\nu} (v \cdot \partial)^3
\delta^d(x) \\
S_\mu S_\nu \partial_\mu \partial_\nu G_0(x) E_1(x;v) &\sim&
- \frac{2}{3(4\pi)^2\ve} S_\mu S_\nu \delta_{\mu\nu} \delta^d(x) \\
S_\mu S_\nu \partial_\mu \partial_\nu G_0(x)v \cdot \partial  E_1(x;v) &\sim&
 \frac{2}{(4\pi)^2\ve} S_\mu S_\nu \delta_{\mu\nu} v \cdot \partial
\delta^d(x) \\
S_\mu S_\nu \partial_\mu \partial_\nu G_1(x) E_0(x;v) &\sim&
 \frac{1}{(4\pi)^2v^2\ve} S_\mu S_\nu \delta_{\mu\nu} \delta^d(x) \\
S_\mu S_\nu \partial_\mu \partial_\nu G_1(x)v \cdot \partial E_0(x;v) &\sim&
- \frac{1}{(4\pi)^2v^2\ve} S_\mu S_\nu \delta_{\mu\nu} v \cdot \partial
\delta^d(x) .
\eeqa
The actual calculation of the divergent part of $\Sigma_2$
is rather tedious. The final result can be expressed in terms of a local
functional $\wh \Sigma_2(y)$ defined in analogy with
(\ref{Si1div}) :
\beq
\Sigma_2^{\rm div}(y,y') = \frac{1}{(4\pi F)^2\ve} \wh\Sigma_2(y)
\delta^d(y-y').
\eeq
Transforming back to Minkowski space, one finds
\beqa
\wh \Sigma_2(y) &=& i\left\{\frac{1}{4} [2(v \cdot u)^2 +
\langle (v \cdot u)^2\rangle] v \cdot \nabla +
\frac{1}{2} v \cdot u(v \cdot \nabla v \cdot u) +
\frac{1}{4} \langle v \cdot u (v \cdot \nabla v \cdot u)\rangle\right\}
\no \\
&& \mbox{} + g_A \left\{- \frac{1}{2} v \cdot u \langle S \cdot u v
\cdot u\rangle + \frac{1}{4} \langle S \cdot u(v \cdot u)^2\rangle
- S^\mu v^\nu [\Gamma_{\mu\nu},v \cdot u] \right\} \no \\
&& \mbox{} + ig^2_A \left\{- \frac{3}{2} (v \cdot \nabla)^3 -
\frac{5}{6} \nabla^\mu \Gamma_{\mu\nu} v^\nu + i \ve^{\mu\nu\rho\sigma}
v_\rho S_\sigma [2\Gamma_{\mu\nu} v \cdot \nabla +
(v \cdot \nabla \Gamma_{\mu\nu})] \right. \no \\
&& \mbox{} - \frac{3}{8} \langle u_\mu (v \nabla u^\mu)\rangle
- \frac{3}{8} (v \cdot \nabla u \cdot u) +
- \frac{9}{32} \langle v \cdot \partial \chi_+\rangle \no \\
&& \mbox{} - \left. \frac{3}{8} [\langle u \cdot u\rangle +
2 u \cdot u + \frac{3}{2} \langle \chi_+\rangle ] v \cdot \nabla\right\}
\no \\
&& \mbox{} + g^3_A \left\{- \frac{1}{2} S \cdot u(v \cdot \nabla)^2 -
2 S^\mu \langle S \cdot u \Gamma_{\mu\nu} \rangle S^\nu -
\frac{1}{2} (v \cdot \nabla S \cdot u) v \cdot \nabla \right. \no \\
&& \mbox{} - \frac{1}{6} ((v \cdot \nabla)^2 S \cdot u) -
\frac{1}{4} u_\mu \langle u^\mu S \cdot u \rangle + \frac{1}{8}
\langle S \cdot u(u \cdot u + \chi_+)\rangle \no \\
&& \mbox{} -  \frac{1}{8} \left.\{ \chi_+,S \cdot u\} + \frac{1}{16}
S \cdot u \langle \chi_+\rangle \right\} \no \\
&& \mbox{} + ig^4_A S_\mu \left\{ [ 2(S \cdot u)^2 - 4 \langle
(S\cdot u)^2\rangle ] v \cdot \nabla + \frac{2}{3}(v \cdot \nabla
S \cdot u) S \cdot u  \right.
\no \\
&& \mbox{} + \left. \frac{4}{3} S \cdot u (v \cdot \nabla S \cdot u)
-  4 \langle S \cdot u(v \cdot \nabla S \cdot u)\rangle
\right\} S^\mu \no \\
&& \mbox{} + g^5_A S_\mu \left\{ \frac{2}{3} (S \cdot u)^3
- \frac{4}{3} \langle (S \cdot u)^3\rangle \right\} S^\mu . \label{S2h}
\eeqa
For the purpose of renormalization, this lengthy expression for
$\wh \Sigma_2$ is not in the most suitable form . For the following
counterterm Lagrangian, powers of the spin matrix are reduced with
the relations
\beq
\{ S^\mu,S^\nu\} = \frac{1}{2} (v^\mu v^\nu - g^{\mu\nu}), \qquad
[S^\mu,S^\nu] = i \ve^{\mu\nu\rho\sigma} v_\rho S_\sigma.
\eeq
In addition to various $SU(2)$ identities, I have also used the curvature
relation
\beq
\Gamma_{\mu\nu} = \frac{1}{4} [u_\mu,u_\nu] - \frac{i}{2} f_{+\mu\nu}.
\eeq

\paragraph{6.} With the conventions of Ref. \cite{GLAP} for separating the
finite part in dimensional regularization, we decompose $Z_{\rm irr}[j,
\rho_v]$ in (\ref{Zirr}) into a finite and a divergent part, both
depending on the arbitrary scale parameter $\mu$:
\beqa
\Sigma_1(y,y') \delta^4(y-y') + \Sigma_2(y,y') &=&
\Sigma_1^{\rm fin}(y,y';\mu) \delta^4(y-y') + \Sigma_2^{\rm fin}(y,y';\mu)
\no \\
&& \mbox{} - \frac{2\Lambda(\mu)}{F^2} \delta^4(y-y')[\wh \Sigma_1(y) +
\wh \Sigma_2(y)] \label{Sidiv} \\
\Lambda(\mu) &=& \frac{\mu^{d-4}}{(4\pi)^2} \left\{ \frac{1}{d-4}
- \frac{1}{2} [\log 4\pi + 1 + \Gamma'(1)] \right\}. \no
\eeqa
The generating functional $Z[j,\rho_v]$ can now be renormalized by
introducing the counterterm Lagrangian
\beq
\cL^{(3)}_{\rm ct}(x) = \frac{1}{(4\pi F)^2} \sum_i B_i \bar N_v(x)
O_i(x) N_v(x) \label{Lct}
\eeq
with dimensionless coupling constants $B_i$ and  field monomials
$O_i(x)$ of $O(p^3)$. In analogy to Eq. (\ref{Sidiv}),
the low--energy constants $B_i$ are decomposed as
\beq
B_i = B_i^r(\mu) + (4\pi)^2 \beta_i \Lambda(\mu). \label{beta}
\eeq
The $\beta_i$ are dimensionless functions of $g_A$ designed to cancel the
divergences of the one--loop functional. In Table~1, the operators
$O_i$ are listed together with the coefficients $\beta_i$.

\begin{table}
\caption{Counterterms and their $\beta$--functions as defined in
Eqs. (\protect\ref{Lct}, \protect\ref{beta})}
$$
\begin{tabular}{|r|c|c|} \hline
i  & $O_i$ & $\beta_i$ \\ \hline
1  & $i[u_\mu,v \cdot \nabla u^\mu]$ & $g^4_A/8$ \\
2  & $i[u_\mu,\nabla^\mu v \cdot u]$ & $- (1 + 5 g^2_A)/12$ \\
3  & $i[v \cdot u, v \cdot \nabla v \cdot u]$ & $(4 - g^4_A)/8$ \\
4  & $S \cdot u \langle u \cdot u \rangle$ & $g_A (4 - g^4_A)/8$ \\
5  & $u_\mu \langle u^\mu S \cdot u\rangle$ & $g_A (6 - 6 g^2_A
+ g^4_A)/12$ \\
6 & $S \cdot u \langle(v \cdot u)^2\rangle$ & $- g_A (8 - g^4_A)/8$ \\
7  & $v \cdot u \langle S \cdot u v \cdot u\rangle$ & $- g^5_A/12$ \\
8  & $[\chi_-,v \cdot u]$ & $(1 + 5 g^2_A)/24$ \\
9  & $S \cdot u \langle \chi_+\rangle$ & $g_A (4 - g^2_A)/8$ \\
10 & $\nabla^\mu f_{+\mu\nu} v^\nu$ & $- (1 + 5 g^2_A)/6$ \\
11 & $i S^\mu v^\nu [f_{+\mu\nu},v \cdot u]$ & $g_A$ \\
12 & $i v_\lambda \ve^{\lambda\mu\nu\rho} \langle u_\mu u_\nu u_\rho\rangle$
& $- g^3_A (4 + 3 g^2_A)/16$ \\
13 & $v_\lambda \ve^{\lambda\mu\nu\rho} S_\rho \langle (v \cdot \nabla u_\mu)
u_\nu\rangle$ & $ - g^4_A/4$ \\
14 & $v_\lambda \ve^{\lambda\mu\nu\rho} \langle f_{+\mu\nu} u_\rho\rangle$
& $- g^3_A/4 $ \\
15 & $i(v \cdot \nabla)^3$ & $- 3 g^2_A$ \\
16 & $v \cdot \stackrel{\leftarrow}{\nabla} S \cdot u v \cdot \nabla$ &
$g^3_A$ \\
17 & $\langle u \cdot u\rangle iv \cdot \nabla +$ h.c. &
$- 3 g^2_A (4 + 3 g^2_A)/16$ \\
18 & $\langle (v \cdot u)^2 \rangle i v \cdot \nabla +$ h.c. &
$ (8 + 9 g^4_A)/16$ \\
19 & $(v \cdot \nabla S \cdot u) v \cdot \nabla +$ h.c. & $
g^3_A/3$ \\
20 & $\langle \chi_+\rangle i v \cdot \nabla +$ h.c. &
$- 9 g^2_A/16$ \\
21 & $v_\lambda \ve^{\lambda\mu\nu\rho} S_\rho u_\mu u_\nu v \cdot \nabla +$
h.c. & $- g^2_A (4 + g^2_A)/4 $ \\
22 & $i v_\lambda \ve^{\lambda\mu\nu\rho} S_\rho f_{+\mu\nu} v \cdot \nabla
+$ h.c. & $g^2_A$ \\ \hline
\end{tabular}
$$
\end{table}

This completes the renormalization of Green functions to $O(p^3)$.
The sum of the irreducible
one--loop functional $Z_{\rm irr}[j,\rho_v]$ and of the counterterm
functional is finite and scale independent:
\beqa
\lefteqn{Z_{\rm irr}[j,\rho_v] + \frac{1}{(4\pi F)^2} \sum_i B_i
\int d^4x d^4x' d^4y \bar \rho_v(x) A^{-1}_{(1)}(x,y) O_i(y)
A^{-1}_{(1)}(y,x') \rho_v(x') = } \no \\
&=& \int d^4 x d^4x' d^4y d^4y' \bar \rho_v(x) A^{-1}_{(1)}(x,y)
[\Sigma_1^{\rm fin} (y,y';\mu) \delta^4(y-y') +
\Sigma_2^{\rm fin} (y,y';\mu) \no \\
&& \mbox{} + \frac{1}{(4\pi F)^2} \delta^4(y-y') \sum_i B_i^r(\mu)
O_i(y)] A^{-1}_{(1)}(y',x') \rho_v(x').
\eeqa
Since the same is true for the sum of the reducible contributions of
Fig.~2, the total generating functional of $O(p^3)$ has been rendered
finite and scale independent.

The renormalized low--energy constants $B_i^r(\mu)$ are measurable
quantities satisfying the renormalization group equations
\beq
\mu \frac{d}{d\mu} B_i^r(\mu) = - \beta_i
\eeq
implying
\beq
B_i^r(\mu) = B_i^r(\mu_0) - \beta_i \log \frac{\mu}{\mu_0}.
\eeq

The operators in Table~1 constitute a minimal set for renormalizing the
irreducible self--energy functional for off--shell nucleons. As long as
one is interested in Green functions with on--shell nucleons only, the
number of operators in (\ref{Lct}) can be further reduced by invoking the
equations of motion for the nucleons. We shall not pursue this option
here as it requires some additional discussion.

Partial results for one--loop divergences in chiral $SU(2)$ have been
obtained before. In particular, the results in Table~1 agree with those
of Ref. \cite{BKKM}. Some of the coefficients have also been calculated
recently by Park, Min and Rho: the coefficients $c_i$ in
Eq. (43) of their paper \cite{Rho} agree with Table~1, except for
$c_4$ which should be a factor two smaller.

\paragraph{7.} In addition to phenomenological applications for chiral
$SU(2)$, the methods of this paper can be applied and extended in
various ways:
\begin{enumerate}
\item[i)] Chiral $SU(3)$.
\item[ii)] Inclusion of higher baryon states, especially in view of
recent results on the $1/N_C$ expansion for baryons
(see Ref.~\cite{DJM} and references therein).
\item[iii)] Renormalization of the one--loop functional of $O(p^4)$.
\item[iv)] Non--leptonic weak interactions of baryons.
\item[v)] Combination of heavy quark and chiral symmetries.
\end{enumerate}

\vfill
\section*{Acknowledgements}
I am indebted to J\"urg Gasser, Heiri Leutwyler and Helmut Neufeld for
helpful discussions and useful comments.

\section*{Figure Captions}
\begin{description}
\item[Fig. 1:] Irreducible one--loop diagrams. The full (dashed) lines
denote the nucleon (meson) propagators. The double lines indicate that
the propagators (as well as the vertices) have the full tree--level
structure attached to them as functionals of the external fields.
\item[Fig. 2:] Reducible diagrams of $O(p^3)$. The cross denotes
counterterms from the meson Lagrangian $\cL_4$ of $O(p^4)$.
\end{description}

\newpage

\newcommand{\PL}[3]{{Phys. Lett.}        {#1} {(19#2)} {#3}}
\newcommand{\PRL}[3]{{Phys. Rev. Lett.} {#1} {(19#2)} {#3}}
\newcommand{\PR}[3]{{Phys. Rev.}        {#1} {(19#2)} {#3}}
\newcommand{\NP}[3]{{Nucl. Phys.}        {#1} {(19#2)} {#3}}

\end{document}